\relax
\documentstyle [12pt]{article}
\begin {document}
\bibliographystyle {plain}
%\tableofcontents

\title{\bf  Scaling Properties of the Two-Chain Model}
\author {A.A.Nersesyan, A.Luther and F.V.Kusmartsev}
\maketitle
\begin {verse}
$Institute~ of~ Theoretical Physics,~
Chalmers~ University~ of~ Technology,$\\ $S-41296,~
Goteborg,~ Sweden$  \footnote { Mailing address\\
e-mail: ners@fy.chalmers.se }\\
$and$\\
$Institute~ of~ Physics,~
Georgian~ Academy~ of~ Sciences,$
\\
$Tamarashvili~ 6,~ 380077,~ Tbilisi, ~Georgia$\\
\end{verse}
\author{A.Luther}
\begin {verse}
$NORDITA,~ Blegdamsvej ~17,~ DK-2100~ Copenhagen ~0,~ Denmark$\\
\end{verse}
\author{F.V.Kusmartsev}
\begin {verse}
$L.D.Landau~ Institute~ for~ Theoretical~ Physics$\\
$Moscow,~ 117940~, GSP-1,~ Kosygin~ 2,~V-334,~ Russia$\\
$and$\\
$Department~ of~ Theoretical~ Physics$,\\
$Oulu~ University,~ Linnanmaa,~ 90570,~ Oulu, ~Finland$
\end{verse}

\begin{abstract}
\par
Scaling properties of a self-dual field-theoretical model, describing two
weakly coupled
spinless Luttinger chains, are studied.
A crossover to a sine-Gordon massive phase, with strongly developed
two-particle
interchain correlations, is described. It is argued that, in a wide range of
the in-chain interaction, renormalization of the interchain hopping amplitude
is determined by the Luttinger liquid effects.
\end{abstract}
\newpage
\sloppy
\par
1. Anderson suggested that in an array of weakly coupled  chains a
single-fermion
interchain hopping ($t_{\perp}$) may be irrelevant (Anderson's confinement)
\cite {And}, implying the possibility of a Luttinger liquid
(LL) behavior of
higher-dimensional ($d > 1$) strongly correlated electron systems \cite {And1}.
The issue of Anderson's confinement, which is indeed a rather complicated
problem even for  chains with spinless fermions, has attracted
much interest [1, 3 - 8].
\par
Two opposite tendencies, associated with single-particle and two-particle
correlations,
characterize the low-energy behavior of a multi-chain system.
On increasing the in-chain interaction $g$, the renormalized
single particle hopping amplitude,
being a measure of coherent delocalization of the particles between the
chains, gets
gradually suppressed due to infrared catastrophe \cite {KLN}.
In the spinful case this suppression is
also contributed by the charge-spin separation \cite {And}.
At least in continuum models with a linear spectrum, such a tendency
eventually leads to deconfinement-confinement transitions when the interaction
becomes large enough [3,5,6].
On the other hand, depending on the sign of $g$, either particle-hole ($ph$) or
particle-particle ($pp$) interchain hoppings are generated in the second
order in $t_{\perp}$ [9,10]. These two-particle processes,
with amplitudes increasing
with $|g|$, inevitably drive the system away from the LL fixed point
to strong-coupling, massive phases characterized by density-wave or pairing
fluctuations, strongly correlated in the transverse direction [4 - 6].
\par
Thus, the problem one faces when studying scaling properties of the two-chain
system
is the interplay between single-particle interchain hopping
and two-particle correlations, described by two relevant operators with
different critical dimensions. Understanding of this interplay is especially
important
at such values of $|g|$, when the two-particle hopping becomes the most
relevant
perturbation. It is not $a~ priori$ clear what is the role of
a large mass gap, generated by
two-particle processes, in the renormalization of $t_{\perp}$.
It is well known that such a mass gap could give rise to a finite threshold
for a perturbation having the same scaling dimension as that of $t_{\perp}$.
Such a situation occurs
in 1d Fermi systems with attractive interaction in a magnetic field \cite
{JN}, or in "commensurate-incommensurate" transitions \cite {PT}, described
by the sine-Gordon model with a finite density of topological charge.
Therefore, a question arises: Is it the LL "$infrared ~catastrophe$" or the
two-particle mass gap which mostly determines renormalization of $t_{\perp}$ ?
Is there any finite threshold for $t_{\perp}$, or not ?
\par
Motivated by these questions, we analyze scaling
properties of a self-dual field-theoretical model, recently proposed \cite
{KLN}
to describe two weakly coupled spinless Luttinger chains. The model naturally
incorporates both the development of single-fermion confinement and interchain
pair coherence. This was demonstrated in \cite {KLN} by using an equivalent
representation of the theory
in terms of a 2d Coulomb gas of charge-monopole composites. We show that, on
decreasing the energy scale, the two-chain system scales down to a
non-Luttinger,
massive phase with strong interchain correlations, effectively described
by a sine-Gordon model.
At weak interaction, when $t_{\perp}$ is the most relevant perturbation, such a
crossover occurs via formation of intermediate two-band LL, with a finite
Zeeman-like interband splitting.
Among possible ground states of the two-chain system, we indicate the orbital
aniferromagnet phase, previously discussed by two of us [13]. Specifically,
this
type of ordering occurs due to interchain forward scattering.
\par
At larger interaction, when the two-particle hopping dominates over the
single-fermion
one, the two-band LL regime is absent. Yet, we find no threshold for
$t_{\perp}$ and argue that $t_{\perp}$ is mostly renormalized by pure LL
screening
effects. This conclusion is supported by using a
relationship with spin-chain models.
\\
\par
2. We consider the following model of two spinless Luttinger chains with a
weak single-particle interchain hopping:
$$
H = \sum_{\mu} \int dx [-iv_{F}(\psi^{+}_{1\mu}\partial_{x}\psi_{1\mu} -
\psi^{+}_{2\mu}\partial_{x}\psi_{2\mu}) +
\pi v_{F} ( g \rho_{1\mu}\rho_{2\mu} + g^{'} \rho_{1\mu} \rho_{2,-\mu})]
$$
\begin{equation}
+ t_{\perp}\sum_{\mu}\int dx (\psi^{+}_{1\mu}\psi_{1,-\mu} +
\psi^{+}_{2\mu}\psi_{2,-\mu})
\end{equation}
where $v_{F}$ is the Fermi velocity, $\mu = \pm 1$ is the chain index,
$\psi_{1\mu}$ and $\psi_{2\mu}$
are the Fermi fields for right-movers and left-movers, respectively, and
$\rho_{j\mu}(x) = ~:\psi^{+}_{j\mu}(x)\psi_{j\mu}(x):$ are the density
operators.
$g$ and $g^{'}$are
dimensionless coupling constants characterizing the in-chain and interchain
forward
scattering, respectively.
Using the standard Abelian bosonization, we transform the original problem of
Fermi fields to a purely bosonic one,
$
H_{B} = H^{0}_{\rho} + H_{\sigma}
$
where $H^{0}_{\rho}$ is a Tomonaga-Luttinger Hamiltonian
describing gapless excitations of the total (or "charge")
density,
$\rho_{j} = (1/\sqrt{2}) \sum_{\mu} \rho_{j\mu}$,
which decouple from the rest of the spectrum. All nontrivial effects caused by
the interchain hopping $t_{\perp}$ are incorporated in the "spin" part
of the Hamiltonian $H_{\sigma} = H^{0}_{\sigma} + H_{\perp}$ which deals with
relative degrees
of freedom, $\sigma_{j} = (1/\sqrt{2}) \sum_{\mu} \mu \rho_{j\mu}$.
The coupling constants emerging in the $\rho$- amd $\sigma$-channels are,
respectively, $g_{\rho, \sigma} = g \pm g^{'}$.
As shown in \cite {KLN}, $H_{\sigma}$ is given by the following
field-theoretical model
\begin{equation}
H_{\sigma} = \int dx [\frac{u}{2}\ (P^{2} + (\partial_{x}\phi)^{2}) +
\frac{2t_{\perp}}{\pi\alpha} cos(\frac{1}{2} \gamma\phi)
cos(\frac{1}{2} \tilde{\gamma} \tilde{\phi})]
\end{equation}
Here $\phi(x)$ and $P(x)$ are a scalar field and its conjugate momentum,
respectively, with a canonical commutation relation
$
[\phi(x,t), P(x^{'},t)] = i \delta (x - x^{'}),
{}~~ u = v_{F} (1 - \frac{1}{4}g^{2}_{\sigma})^{1/2}$
is the renormalized velocity,
$\tilde{\phi}(x)$ is a field dual to $\phi(x)$, defined as
$
\partial_{x} \tilde{\phi}(x) = P(x),
$
and
\begin{equation}
\frac{\gamma^{2}}{8\pi}\ = \frac{8\pi}{\tilde{\gamma}}
\equiv K = (1 - g_{\sigma}/2)^{1/2}
(1 + g_{\sigma}/2)^{-1/2}
\end{equation}
The fields $\phi$ and $\tilde{\phi}$ are related to the "spin" density and
current:
$
\sigma_{1} + \sigma_{2} = (\gamma /\sqrt{8} \pi) \partial_{x} \phi, ~~
\sigma_{1} - \sigma_{2} = - (\tilde{\gamma} /\sqrt{8} \pi) \partial_{x}
\tilde{\phi}.~$
The model (2) is self-dual: $H_{\sigma}$ is invariant under transformations
$\phi \leftrightarrow \tilde{\phi} ~(g_{\sigma} \rightarrow - g_{\sigma}), ~
 \gamma \leftrightarrow \tilde{\gamma}$.
\par
The perturbation term in (2) has critical dimension
$\Delta = \frac{1}{2}(K + \tilde{K})$ \cite {Wen}
and, hence, is relevant when $\Delta < 2$. This inequality corresponds to the
interval $K_{-} < K < K_{+}$,
where the points $K_{\pm} = 2 \pm \sqrt{3}$ mark the boundaries between the
confinement and deconfinement phases \cite {KLN}. As follows from the Coulomb
gas
representation of the model (2) \cite {KLN}, the single-particle confinement
originates from pairing of the charge-monopole composites with zero total
"electric" and "magnetic" charges. On the other hand, binding of composites in
pairs
with total "magnetic" or "electric" charge $\pm 2$ is described by two
operators
\begin{equation}
O_{ph} = (\psi^{+}_{1\mu} \psi_{1,-\mu})(\psi^{+}_{2,-\mu}\psi_{2\mu})
\rightarrow - \frac{1}{2(\pi\alpha)^{2}} \cos \gamma \phi,
\end{equation}
\begin{equation}
O_{pp} = (\psi^{+}_{1\mu} \psi_{1,-\mu})(\psi^{+}_{2 \mu}\psi_{2,-\mu})
\rightarrow - \frac{1}{2(\pi\alpha)^{2}} \cos \tilde{\gamma} \tilde{\phi}
\end{equation}
associated with interchain particle-hole ( $ph$) and
particle-particle ($pp$) hoppings, respectively. These two-particle processes,
although
absent in the initial Hamiltonian (2), are generated upon renormalization.
The critical dimensions of operators (4) and (5) are,
respectively, $2K$ and $1/2K$. So, at any $K \neq 1$ ($g_{\sigma} \neq 0$),
one of these operators is always relevant. Therefore, the two-particle
hopping processes must be included into renormalization scheme [10,14].
\par
At the $l$-th step of the renormalization procedure ($l$ can be identified as
a continuous logarithmic variable $l = ln(\alpha/\alpha_{0})$
or $l = ln (\Lambda/|\omega|)$), the Hamiltonian density takes the form
$$
H_{l}(x) = \frac{u_{l}}{2} \left( P^{2} + (\partial_{x} \phi)^{2} \right)
+ \frac{4 u_{l} \tau_{l}}{\alpha^{2}} \cos (\frac{1}{2} \gamma_{l} \phi)
\cos (\frac{1}{2} \tilde{\gamma}_{l} \tilde{\phi})
$$
\begin{equation}
- \frac{2 \pi u_{l}}{(2 \pi \alpha)^{2}} \left ( G_{l} \cos (\gamma_{l} \phi)
+ \tilde{G}_{l} \cos (\tilde{\gamma}_{l} \tilde{\phi}) \right )
\end{equation}
where $\gamma_{l} = 8 \pi/\tilde{\gamma}_{l}$
is given by formula (3), in which $g_{\sigma}$ is replaced by $z_{l}$;
$
\tau_{l} = t_{\perp}(l) \alpha / 2 \pi u_{l}
$
is a renormalized dimensionless single-fermion hopping amplitude.
\par
The parameters of the Hamiltonian $H_{l}$ are determined from the following
renormalization-group equations
\footnote{
There is one more equation,
$
\zeta^{'}_{l} = - 4 \pi^{2} (\Delta_{l} + 1) \tau^{2}_{l},
$
which describes renormalization of the velocity:
$
u_{l} \rightarrow u_{l} (1 - \zeta_{l}).
$
This renormalization reduces
to rescaling of $K$ and $\tilde{K}$:
$K \rightarrow K (1 - \zeta^{2})^{-1/2}, ~
\tilde{K} \rightarrow \tilde{K} (1 - \zeta^{2})^{1/2},
$
which will be assumed everywhere below.}
\begin{equation}
\tau^{'}_{l} = (2 - \Delta_{l}) \tau_{l}
\end{equation}
\begin{equation}
G^{'}_{l} = 2(1 - K_{l}) G_{l} + (K_{l} - \tilde{K}_{l}) \tau^{2}_{l}
\end{equation}
\begin{equation}
\tilde{G}^{'}_{l} = 2(1 - \tilde{K}_{l}) \tilde{G}_{l} +
(\tilde{K}_{l} - K_{l}) \tau^{2}_{l}
\end{equation}
\begin{equation}
(ln~K)^{'} = \frac{1}{2} (\tilde{K}_{l} \tilde{G}^{2}_{l} -
K_{l} G^{2}_{l} )
\end{equation}
with initial conditions
$
u_{0} = v_{F},~~K_{0} = K, ~~G_{0} = \tilde{G}_{0} = 0,~
$
and $ \tilde{K}_{l} K_{l} = 1.~$
The upper prime indicates the derivative $\partial/\partial l$.
Equations (8) and (9) were given by Yakovenko \cite {Yak}. However, he did not
consider
renormalization of the coupling constant $g_{\sigma}$, Eq.(10), which,
as will be shown below,
is crucial for a weak interaction, when a crossover from a two-chain Luttinger
liquid
to a strong-coupling low-energy regime is developed.
\\
\par
3. We consider first the case of a weak interaction, when one can set
\begin{equation}
K_{l} \simeq 1 - \frac{1}{2} z_{l},~~ \tilde{K}_{l} \simeq 1 + \frac{1}{2}
z_{l},~~
\Delta_{l} = 1 + 0(z^{2}_{l}).
\end{equation}
Solving Eq.(7),
one determines the value of the variable
$
l_{0} = ln (1/\tau_{0}) = ln (\Lambda/\omega_{0}),
$
at which $\tau_{l}$ becomes $\sim 1$. This introduces a new
energy scale to the problem - a Zeeman-like splitting of two degenerate bands:
$
\omega_{0} \simeq \Lambda \tau_{0} \sim t_{\perp}.
$
\par
In the region
$l \leq l_{0}$, or $\omega_{0} < |\omega| < \Lambda$,
$z_{l}$ is not renormalized (up to second-order corrections in
$g_{\sigma}$), $z(l_{0}) = g_{\sigma}$,
while the charges $G_{l}$ and $\tilde{G}_{l}$ remain small:
\begin{equation}
G_{l} = -\tilde{G}_{l} = - \frac{1}{2} g_{\sigma} \tau^{2}_{0} ( e^{2l} -
e^{gl}) \simeq
- \frac{1}{2} g_{\sigma} \tau^{2}_{l}
\end{equation}
To solve the renormalization-group equations in the low-energy region
$l \geq l_{0}$, we make use of a typical two-cutoff scaling prescription:
understanding that
at $l \sim l_{0}$ renormalization of $t_{\perp}(l)$ is stopped, we
choose the normalization condition $\tau_{l_{0}} = 1$ and
drop the $\tau^{2}$-terms in r.h.s. of Eqs.(8) and (9).
Then we arrive at the following
set of equations in the region $l \geq l_{0}$:
\begin{equation}
z^{'} = G^{2} - \tilde{G}^{2}, ~~~~
G^{'} = z G, ~~~~
\tilde{G}^{'} = - z \tilde{G}
\end{equation}
with the boundary conditions
$
z(l_{0}) = g_{\sigma}, ~~ G(l_{0}) = - \tilde{G}(l_{0}) = - \frac{1}{2}
g_{\sigma}.
$
\par
Eqs.(13) describe a $Z_{4}$-symmetric
model. Its 2d classical version is a planar XY model in the presence of
4-fold degenerate symmetry breaking field [15,16].
Such a model is also
equivalent to a version of the backscattering model of 1d fermions
\cite {LE} extended to
include spin-nonconserving processes \cite {Truong}. We recognize a
field-theoretical version of this model in (6) when $t_{\perp}$ is dropped.
At $l = l_{0}$ this model reads
\begin{equation}
H_{l_{0}} = \frac{v_{F}}{2} \left( P^{2} + (\partial_{x}\phi)^{2} \right)
+ \frac{\pi v_{F} g_{\sigma}}{(2 \pi \alpha)^{2}} (\cos \gamma \phi - \cos
\tilde{\gamma} \tilde{\phi})
\end{equation}
\par
Note that, in Eq.(14) amplitudes of the cosines coincide. This reflects the
"hidden" U(1) symmetry of the model (14) which
 brings it to the sine-Gordon (SG) universality
class, corresponding to a critical plane of the $Z_{4}$-model \cite {Giam}.
This is directly seen from the solution of Eqs.(13).
Introducing linear combinations
$
G_{\pm} = G \pm \tilde{G}
$
and taking into account boundary conditions at $l = l_{0}$, one
finds that, at all $l > l_{0}$, $z = - G_{-}$, thus
reducing Eqs.(13)
to a pair of Kosterlitz-Thouless equations for the SG model:
$
{}~~G^{'}_{+} = - G^{2}_{-}, ~~~ G^{'}_{-} = - G_{+} G_{-}.
$
The conditions
$
G_{+}(l_{0}) = 0, ~~ G_{-}(l_{0}) = - g_{\sigma}
$
define a scaling trajectory with the initial point lying on the vertical
axis of the standard Kosterlitz-Thouless phase plane, $(G_{+}, |G_{-}|)$.
Therefore, for any sign
of $g_{\sigma}$, a strong-coupling regime develops in the infrared limit.
The dynamically generated mass gap
\begin{equation}
M \simeq t_{\perp} \exp ( - \pi / 2|g_{\sigma}| ) << t_{\perp}
\end{equation}
signals the onset of strong interchain two-particle correlations.
The fact that the mass does not depends on the sign of  $g_{\sigma}$
is related to self-dual symmetry of the model.
\\
\par
4. Let us consider our model in the new basis which
is built up from symmetric and antisymmetric states
$
\Psi_{j,\pm} = (\pm \psi_{j,+} + \psi_{j,-})/\sqrt{2}.
$
In the absence of the interchain hopping,
 two  bands associated with these states are degenerate. The degeneracy is
removed
by interchain hopping, and the resulting Zeeman-like splitting of the two bands
is a measure of coherent delocalization of fermions between the chains.
\par
When the chain index is formally treated as a spin-1/2 variable, the interchain
hopping term of the Hamiltonian turns to Zeeman interaction with a magnetic
field
$2t_{\perp}$ along the $x$-axis. In the new basis of states, which is
obtained by a $\pi/2$ - rotation about the $y$-axis in "spin" space, the
magnetic
field is oriented along the $z$-axis, and the splitting of the bands is
determined
by "magnetization" $<S^{z}>$.
\par
The $\pi/2$ - rotation of the basis has no effect on the "charge" degrees of
freedom.
However, it produces the following changes in the $\sigma$-channel:
\begin{equation}
O_{ph, pp} \rightarrow \sigma_{1} \sigma_{2} \pm \frac {1}{2} (O_{ph} -
O_{pp}), ~~
\sigma_{1} \sigma_{2} \rightarrow \frac {1}{2} (O_{ph} + O_{pp})
\end{equation}
Quantities in the r.h.s. of (16) now have a different meaning: $\sigma_{j}$ is
a
difference between the particle densities in each of the bands, while the
operators
$O_{ph}$ and $O_{pp}$ describe $interband$ pair-hopping processes
\par
Making use of (16), we transform the two-chain Hamiltonian (6) to a $two-band$
one:
\begin{equation}
H^{'}_{l} = \frac{u_{l}}{2} \left( \Pi^{2} + (\partial_{x} \chi)^{2} \right)
+ \frac{\beta_{l} t_{l}}{2 \pi} \partial_{x} \chi +
\frac{2 \pi u_{l}}{(2 \pi \alpha)^{2}} \left( z_{\perp}(l) \cos (\beta_{l}\chi)
+ z_{f}(l) \cos (\tilde{\beta}_{l} \tilde{\chi}) \right )
\end{equation}
where
$$
\frac{\beta^{2}_{l}}{8\pi} = \frac {8\pi}{\tilde{\beta}^{2}_{l}}
= \left( \frac{1 - \frac{z_{\parallel}(l)}{2}}{1 + \frac {z_{\parallel}(l)}{2}}
\right)^{1/2}, ~~~z_{\parallel}(l) = G_{l} + \tilde{G}_{l}
$$
\begin{equation}
z_{\perp}(l) = \frac{1}{2} (z_{l} + G_{l} - \tilde{G}_{l}), ~~~
z_{f}(l) = \frac{1}{2} (z_{l} - G_{l} + \tilde{G}_{l})
\end{equation}
with initial conditions
\begin{equation}
z_{\parallel}(0) = 0, ~~~ z_{\perp}(0) = z_{f}(0) = \frac{1}{2} g.
\end{equation}
\par
In Eq. (17) we recognize the above mentioned U(1) symmetric
1d Fermi-gas model
with backscattering ($z_{\perp}$) and spin-nonconserving processes ($z_{f}$)
in a magnetic field
along the $z$-axis \cite {Giam}. At $l \leq l_{0}$ the effective couplings
equal
\begin{equation}
z_{\parallel}(l) = 0, ~~~ z_{\perp}(l) = \frac{1}{2} g_{\sigma} (1 -
\tau_{l}^{2}),
{}~~~z_{f}(l) = \frac{1}{2} g_{\sigma} (1 + \tau_{l}^{2}).
\end{equation}
 One sees that, at energies $|\omega| \sim t_{\perp}$ ($l \sim l_{0}$),
splitting of the two bands ($\tau_{l}$)
tends to suppress
backscattering (or interband $ph$ transitions, $z_{\perp}(l)$), which is due to
nonconservation of the total momentum.  The value of $l_0$ in
the two-cutoff scaling is determined by condition $z_{\perp}(l) = 0$.
\par
The suppression of the $z_{\perp}$-term breaks the tiny balance between the
amplitudes of the two cosines, which preserves a Tomonaga-Luttinger
weak-coupling
behavior at $t_{\perp} = 0$, and results in a crossover to a SG model
\begin{equation}
H^{'}_{l_{0}} = \frac{v_{F}}{2} \left ( \tilde{\Pi}^{2} +
(\partial_{x} \tilde{\chi})^{2} \right ) +
\sqrt{\frac{2}{\pi}} t_{l_{0}} \tilde{\Pi} -
\frac{2 \pi v_{F}}{(2 \pi \alpha)^{2}} g_{\sigma} \cos (\tilde{\beta}_{0}
\tilde{\chi}),
\end{equation}
where $\tilde{\beta}_{0} = \beta_{0} = \sqrt{8 \pi}$.
This model is written in terms of the dual field $\tilde{\chi}$,
 using the relationship
$\partial_{x} \chi = \tilde{\Pi}$.
Notice that $t_{l_{0}}$ is coupled to the momentum of the sine-Gordon field
$\tilde{\chi}$. Then, by Galilean invariance,
the term linear in $\tilde{\Pi}$ is eliminated by a shift
$
\tilde{\Pi} = \tilde{\Pi}^{'} - (\sqrt{2} t_{l_{0}}/\sqrt{\pi} v_{F})
$
and does not influence the scaling properties of the system when the energy is
further
decreased in the region $l > l_{0}$. The remaining SG model is precisely the
rotated
version of $H_{l_{0}}$ in (14).
Since $<\tilde{\Pi}^{'}> = 0$, a response to the $t_{\perp}$-perturbation,
being just the magnetization of the "rotated" (the two-band) model, equals
\begin{equation}
<S^{z}>_{l=l_{0}} = - \sqrt{\frac{2}{\pi}}
<\partial_{x} \chi>_{l=l_{0}}
= - \sqrt{\frac{2}{\pi}}<\tilde{\Pi}>_{l=l_{0}}
= \frac{2t_{l_{0}}}{\pi v_{F}}
\end{equation}
One has to take into account scale transformation of the coordinate $x$:
$
\left ( \partial_{x} \chi \right )_{l=0} =
e^{-l_{0}} \left ( \partial_{x} \chi \right )_{l=l_{0}}.
$
Then one finds that $t_{l_{0}}$ should be changed by $t_{\perp}$ in (22) when
true response $<S^{z}>_{l=0}$ is considered.
\par
Since the $0(g^{2})$-correction to $\Delta$ was omitted, the obtained
 two-band splitting coincides with that for the noninteracting
fermions. It can be easily checked that keeping such a correction
is still within
 the main accuracy to which the above renormalization treatment of the
case $|g_{\sigma}| << 1$ is valid. Then we obtain
\begin{equation}
<S^{z}> = \frac{2 t_{\perp}}{\pi v_{F}}
\left ( 1 - \frac{1}{8}g_{\sigma}^{2} ln \frac{1}{\tau_{0}} + ... \right ).
\end{equation}
\par
 The strong-coupling regime developing in the
low-energy region $(|\omega| \leq \omega_{0})$ is related to ordering of the
$dual$ field $\tilde{\chi}$. In terms of the original model (6), this
includes both possibilities for ordering of the field $\phi$ at $g_{\sigma} >
0$
($ph$ hopping) and the dual field $\tilde{\phi}$ at $g_{\sigma} < 0$
($pp$ hopping).
The field $\chi$ remains disordered and,
therefore, is free to adjust itself to a finite "magnetic field" $2t_{\perp}$,
thus acquiring a nonzero expectation value $<\partial_{x} \chi>$.
\\
\par
5. Let us clarify the symmetry of the ground state of the system, characterized
by
ordering of the field $\tilde{\chi}$. As follows from (21), its vacuum
expectation value
depends on the sign of $g_{\sigma}$: $<\tilde{\chi}> = 0$ at $g_{\sigma} > 0$,
and
$<\tilde{\chi}> = \pi/2\tilde{\beta}_{0}$ at $g_{\sigma} < 0$. Therefore, when
bosonizing various bilinears in Fermi fields to construct order parameters in a
$\rho-\sigma$ factorized form, only those operators proportional to
$\cos(\tilde{\beta}_{0}\tilde{\chi}/2)$ or
$\sin(\tilde{\beta}_{0}\tilde{\chi}/2)$
should be considered.
\par
There are two pairs of competing states, with order parameters:
\begin{equation}
O_{CDW} = \sum_{\sigma} \Psi^{+}_{1\sigma} \Psi_{2,-\sigma}
= \sum_{\mu} \mu \psi^{+}_{1\mu} \psi_{2\mu} \sim
\exp(-i\sqrt{2\pi K_{\rho}} \chi_{\rho}) \cos \frac{1}{2}
\tilde{\beta}_{0}\tilde{\chi},
\end{equation}
\begin{equation}
O_{S1} = \sum_{\sigma} \Psi_{1\sigma} \Psi_{2\sigma}
= \sum_{\mu} \psi_{1\mu} \psi_{2\mu} \sim
\exp(-i\sqrt{2\pi \tilde{K}_{\rho}} \tilde{\chi}_{\rho}) \cos \frac{1}{2}
\tilde{\beta}_{0}\tilde{\chi},
\end{equation}
and
\begin{equation}
O_{S2} = i \sum_{\sigma} \sigma \Psi_{1\sigma} \Psi_{2\sigma}
= i \sum_{\mu} \psi_{1\mu} \psi_{2,-\mu} \sim
\exp(-i\sqrt{2\pi \tilde{K}_{\rho}} \tilde{\chi}_{\rho}) \sin \frac{1}{2}
\tilde{\beta}_{0}\tilde{\chi},
\end{equation}
\begin{equation}
O_{OAF} = i \sum_{\sigma} \sigma \Psi^{+}_{1\sigma} \Psi_{2,-\sigma}
= i \sum_{\mu} \mu \psi^{+}_{1\mu} \psi_{2,-\mu} \sim
\exp(-i\sqrt{2\pi K_{\rho}} \chi_{\rho}) \sin \frac{1}{2}
\tilde{\beta}_{0}\tilde{\chi}
\end{equation}
Here the field $\chi_{\rho}$ and its dual field $\tilde{\chi}_{\rho}$ describe
gapless fluctuations in the $\rho$-channel, while
$K_{\rho} \simeq 1 - \frac{1}{2}g_{\rho}$ and
$\tilde{K}_{\rho} \simeq 1 + \frac{1}{2}g_{\rho} ~~(|g_{\rho}| << 1)$
are critical exponents which determine power-law asymptotics of the
corresponding
correlation functions. The order parameters (24) - (27) are given both in the
two-band and two-chain representations. These are two 1d charge-density waves
with "antiferromagnetic" interchain ordering (CDW), the in-chain (S1) and
interchain (S2) Cooper pairings, and orbital antiferromagnetic state (OAF),
characterized by nonzero local currents circulating in the two-chain system
with
$2k_{F}$-periodicity \cite {NL}.
The operator (27) actually
describes only transverse (interchain) currents. The alternating longitudinal
(in-chain) currents are easily shown to be proportional to
$(\partial_{x} \chi) O_{OAF} \sim (t_{\perp}/t_{\parallel}) O_{OAF}$,
as it should be due to the current conservation law.
All operators,  other than (24)-(27), characterize states with short-ranged
correlations, exponentially decaying at distances $|x| > 1/M$.
\par
The dominant ordering among CDW and S1 at $g_{\sigma} > 0$ and among S2 and OAF
at $g_{\sigma} < 0$ is determined by the sign of $g_{\rho}$. The resulting
phase
diagram is shown in Fig.1. It reflects the dual symmetry of the model: States
OAF and S2 are dual to CDW and S1, respectively, and transform to each other
under $g_{\sigma} \rightarrow -g_{\sigma}$.
Notice the important role of
interchain forward
scattering $g^{'}$ which, in addition to already known states [10,5],
 opens new possibilities for ordering of the two-chain system,
such as the OAF state \cite {NL}.
More exotic phases, $e.g.$ spin nematic,
are expected in the case of spin-1/2 particles.
\\
\par
6. At weak interaction the mass gap $M$, Eq.(15), characterizing strong
interchain
two-particle correlations, is exponentially small compared to the interband
splitting $\omega_{0}$. The crossover from a two-chain LL
($\omega_{0} << |\omega| << \Lambda $) to a strong-coupling, quasi-ordered
phase
($|\omega| \leq M$) occurs via formation of a two-$band$ LL
($M << |\omega| << \omega_{0}$). On increasing $g_{\sigma}$, the difference
between $\omega_{0}$ and $M$  decreases, and at large enough interaction the
two-band LL regime no longer exists. In this case,
one of the two-particle hopping amplitudes, $G_{l}$ or $\tilde{G}_{l}$, goes to
strong coupling at such values of $l$, when $\tau_{l}$ is still small [14].
Questions which naturally arise, are: Can a small Zeeman-like splitting of
the two bands survive the presence of a large mass gap M ?
Is the mechanism of suppression of the single-particle interchain hopping,
caused by
LL effects (infrared catastrophe), essentially modified by strongly developed
pair-hopping processes ? These questions cannot be answered by a simple
comparison of
rates of increase of $G_{l}$ (or $\tilde{G}_{l}$) and $\tau^{2}_{l}$ \cite
{Yak},
since such an oversimplified treatment neglects different physical origin of
the two
relevant operators.
\par
Let us assume for definiteness that $g_{\sigma} > 0)$ and consider the case
$ K_{0} < \sqrt{2} - 1 $, when the rate of increase of
$G_{l}$ exceeds that of $\tau_{l}^{2}$. Neglecting
renormalization of $K$, unimportant at large interaction, from Eqs.(7) - (9)
we obtain
\begin{equation}
G_{l} \simeq - \frac{K_{0} - \tilde{K}_{0}}{2 + K_{0} - \tilde{K}_{0}}
\tau^{2}_{0} e^{2(1-K_{0})l}, ~~~~
\tilde{G}_{l} \simeq \frac{\tilde{K}_{0} - K_{0}}{2 + \tilde{K}_{0} - K_{0}}
\tau^{2}_{l}
\end{equation}
We see that $G_{l}$ is increasing in such a way, as if its
initial value were $\sim \tau^{2}_{0}$. The mass gap is then estimated as
\begin{equation}
M \simeq \Lambda~ [C(K_{0}) \tau^{2}_{0}]^{\frac{1}{2(1-K_{0})}}
\end{equation}
where $C(K_{0})$ is a positive (nonuniversal) number.
\par
Note that $G_{l} < 0$. Moreover, after a few renormalizations, $|G_{l}|
>> \tilde{G}_{l}$. This allows to estimate parameters of the "rotated"
Hamiltonian
(17):
\begin{equation}
z_{\parallel}(l) \simeq - |G_{l}|,~~~
z_{\perp}(l) \simeq \frac{1}{2} (g_{l} - |G_{l}|), ~~~
z_{f}(l) \simeq \frac{1}{2} (g_{l} + |G_{l}|)
\end{equation}
Then we find that $\beta^{2}_{l}$ defined by (18) is larger that $8\pi$ and
increases upon renormalization. This indicates irrelevance of $\cos
\beta_{l}\chi$
and relevance of $\cos \tilde{\beta}_{l} \tilde{\chi}$ in (17). This is also in
agreement with scaling behavior of the effective amplitudes $z_{\perp}(l)$ and
$z_{f}(l)$: the former increases upon renormalization, while the latter
decreases.
\par
These observations allow us to conclude that, even in the case when
the interaction
is not small, the model (17) scales towards a strong-coupling regime
characterized
by ordering of the dual field $\tilde{\chi}$. Then, following arguments used
in sec.4, we expect the average magnetization $<S^{z}>$
to be basically determined by scaling equation (7); therefore the interband
splitting
is given by
\begin{equation}
\omega_{0} \simeq \Lambda \tau_{0}^{\frac{1}{2 - \Delta_{0}}}
\end{equation}
\par
Of course, such estimations are valid up to some nonuniversal renormalizations,
inevitable when interaction is not small. Nevertheless, it seems quite
convincing
that a large mass gap,
indicating strong interchain pair coherence, coexists with a finite interband
splitting. The relevance of the single-fermion hopping is then mostly
determined by the
Luttinger critical dimension of the $t_{\perp}$-perturbation in the original
two-chain Hamiltonian (2). Therefore the statement, made in \cite {KLN}
concerning
the Kosterlitz-Thouless-like nature of confinement-deconfinement transitions,
is basically correct.
\par
The fact that the "rotated" model (17) always shows ordering of the dual field
$\tilde{\chi}$ is related to self-duality of the original model (2), still
preserved when pair hopping is taken into account. As a counter example,
violating this property,
one could consider a two-chain model with interchain backscattering,
$$
g_{b} \psi^{+}_{1\mu} \psi^{+}_{2,-\mu} \psi_{1,-\mu} \psi_{2\mu}
\sim g_{b} \cos (\gamma \phi)
$$
which changes the initial condition for
$G_{l}$: $G_{0} = g_{b}$.
Using the "rotated" version of the model, Eq.(17), and repeating the same
analysis as that in sec.3 and 4, it can be easily shown that, under certain
conditions
($g_{b} < 0, ~ 0 < g_{\sigma} < |g_{b}|$), the interchain backscattering drives
the
system to a new strong-coupling fixed point.
The latter is characterized by
irrelevance of the dual field $\tilde{\chi}$ and
ordering of the field $\chi$, which describes two, "ferromagnetically"
correlated,
charge-density waves along the chains.
This type of ordering
does tend to suppress the effective single-particle interchain hopping,
since $t_{\perp}$ now is coupled to the density of topologocal charge of the SG
field $\chi$.
Like in the attractive 1d Fermi system in a magnetic field \cite {JN},
a finite interband splitting will be only nonzero, if $t_{\perp}$ exceeds the
corresponding mass gap, with square-root singularities near the threshold.
In such a situation, the confinement-deconfinement transition, taking place on
increasing the amplitude $t_{\perp}$, is of different nature.
\\
\par
7. To support the above conclusion concerning the existence of a finite band
splitting in the presence of a large mass gap, we use a relationship between
the
two-chain model (2) and quantum spin-1/2 chains.
Consider an aniferromagnetic XY spin-1/2 chain in a homogeneous magnetic field
in the basal plane (e.g. along the x-axis)
\begin{equation}
H = J \sum_{n=1}^{N} (S^{x}_{n}S^{x}_{n+1} + S^{y}_{n}S^{y}_{n+1})
+ h \sum_{n=1}^{N} S^{x}_{n}, ~~(J>0)
\end{equation}
and assume that $|h| << J$. The last condition makes it possible to consider
the continuum limit of the model (32). Using the Jordan-Wigner representation
of the spin operators and the standard bosonization procedure \cite {LP},
we arrive at the following field theory:
\begin{equation}
H = \frac{v_{F}}{2}  \int dx \left ( P^{2}(x) + (\partial_{x} \phi)^{2} \right
)
+ \frac{h}{\sqrt{2 \pi \alpha a}} \int dx \cos (\sqrt{4\pi} \phi(x))
\cos (\sqrt{\pi} \tilde{\phi})
\end{equation}
We see that Hamiltonian (33) is a special case of the model (2) corresponding
to
the value $ K = 2 $ or, by duality, $ K = 1/2 $.
At these particular values of $K$ the spin-chain model (32), with $ h \sim
t_{\perp}$,
represents a lattice version of the two-chain model.
\par
Although the exact solution of the spin model (32) is unknown, we
expect a finite homogeneous magnetization along the x-axis to exist in the
ground
state, which is
equivalent to a finite delocalization of fermions between the chains.
At $h = 0$
the ground state of the XY-model is disordered due to continuous U(1) symmetry,
and the excitation spectrum is gapless. The magnetic field,
breaking U(1) down to $Z_{2}$, will lead to the appearance of a
finite magnetization along the x-axis and to the opening of a gap,
related to antiferromagnetic long-range ordering
of the spins along the y-direction.
The spin order parameter is given by the staggered magnetization $(-1)^{n}<
S^{y}_{n}>$,
which, in the continuum limit, reduces to
$ <\sin (\sqrt{\pi} \tilde{\phi}(x))>$ at $ K = 2 $, or
$ <\sin (\sqrt{\pi} \phi(x))>$ at $ K = 1/2 $, the latter two averages being
the
order parameters of the model (6), related to the development of the $ph$ or
$pp$
pair coherence in the two-chain system.
\par
A more detailed study of the correspondence between the two-chain problem
and quantum spin chains will be presented elsewhere.
\\
\par
One of the authors (A.N.) acknowledges financial support from Chalmers
University of Technology. He also would like to thank H.Johannesson,
S.Kivelson, Yu Lu, T.Einarsson and S.Ostlund for helpful discussions
on the two-chain problem.  F.K. is grateful to C.Bourbonnais
for many fruitful discussions.
Two of us (F.K. and A.N.) acknowledge the
generous hospitality of Nordita, where part of this work was done.

\newpage
{\bf FIGURE CAPTION}
\ \\
Fig.1
\ \\
The phase diagram on the plane $(g, g^{'})$. $g$ and $g^{'}$ are the in-chain
and
interchain forward scattering amplitudes, respectively.

\end{document}